\documentclass[aps,12pt,superscriptaddress,amsfonts,amssymb,amsmath]{revtex4-2}
\pdfoutput=1

\usepackage{graphicx}
\usepackage{makeidx}
\usepackage{xcolor}
\usepackage{amsmath}
\usepackage{bm}
\usepackage{amssymb}
\usepackage{pgfplots}
\pgfplotsset{compat=1.18}

\definecolor{reviewerone}{RGB}{0,70,170}
\definecolor{reviewertwo}{RGB}{185,0,0}

\begin{document}

\title{Optical and magnetic penetration depths under a modified electromagnetic coupling in charged bosonic condensates}

\author{F.\ Minotti \footnote{Email address: minotti@df.uba.ar}}
\affiliation{Universidad de Buenos Aires, Facultad de Ciencias Exactas y Naturales, Departamento de F\'{\i}sica, Buenos Aires, Argentina}
\affiliation{CONICET-Universidad de Buenos Aires, Instituto de F\'{\i}sica Interdisciplinaria y Aplicada (INFINA), Buenos Aires, Argentina}

\author{G.\ Modanese \footnote{Email address: giovanni.modanese@unibz.it}}
\affiliation{Free University of Bozen-Bolzano \\ Faculty of Engineering \\ I-39100 Bolzano, Italy}
\date{\today}

\linespread{0.9}

\begin{abstract}
We establish the relation between optical condensate spectral weight and static magnetic screening in a proposed effective theory of charged bosonic condensates with an additional diamagnetic coupling. Starting from an explicit Ginzburg--Landau functional, we derive both the London equation and the transverse low-frequency conductivity. At fixed condensate parameter $q^2n_s/m^\ast$, the modified coupling doubles the inverse-square magnetic screening length and the optical condensate weight, giving $\lambda_{\rm mag}=\lambda_{\rm stiff}=\lambda_L/\sqrt{2}$, where $\lambda_L^{-2}=\mu_0q^2n_s/m^\ast$. Their ratio is therefore unity in both the standard and modified local theories: it tests consistency between measurement methods but cannot distinguish the couplings. A structured comparison of published results for Nb, Pb, YBCO, MgB$_2$, and Ba(Fe,Co)$_2$As$_2$ identifies the roles of sample matching, absolute calibration, nonlocality, and spectral-weight extraction. We formulate the requirements for a discriminating absolute-response experiment and discuss a complementary single-particle spectral prediction. The analysis separates experimentally accessible consistency checks from tests requiring independent information about condensate parameters or a different observable.
\end{abstract}

\maketitle

\section{Introduction}

Magnetic screening and the low-frequency optical condensate weight are two experimentally accessible manifestations of the transverse response of a charged condensate. Comparing them is useful both for assessing electrodynamic models and for checking the consistency of measurements obtained with different probes. Their ability to discriminate between couplings, however, depends on whether the two observables acquire different modifications. A common rescaling of the response cancels from their ratio.

In this work we establish this cancellation for the restricted effective coupling proposed in Ref.~\cite{MinottiModanese2026GaugeCoupling}. Starting from its effective Ginzburg--Landau functional, we derive the magnetic screening equation and the optical conductivity within the same local transverse approximation. The additional diamagnetic term doubles both the inverse-square magnetic length and the condensate spectral weight at fixed $q^2n_s/m^\ast$. Consequently, the operational optical and magnetic lengths remain equal. This identifies a precise limitation of penetration-depth comparisons as coupling tests, while retaining their value as cross-method consistency checks.

The second contribution is a structured assessment of published optical and magnetic measurements. We distinguish independent absolute determinations from calibrated relative measurements and classify comparisons by specimen matching and material-dependent corrections. This assessment specifies what a controlled comparison can establish and which uncertainties must be resolved before an absolute-response test becomes informative. Finally, we identify two routes beyond the ratio: a determination of the response relative to independently justified condensate parameters, and a complementary spectral observable derived from the modified single-particle Hamiltonian.

From the theoretical point of view, we recall that local gauge invariance (LGI) and minimal coupling are cornerstone principles of modern electrodynamics and of the Standard Model.  In the familiar fermionic case (Dirac QED), replacing $\partial_\mu\to\partial_\mu+i(q/\hbar)A_\mu$ produces an interaction term of the form $-J_\mu A^\mu$ with $J^\mu=q\bar\psi\gamma^\mu\psi$, which coincides with the conserved Noether current associated with global phase symmetry.

For scalar fields, minimal coupling also works and produces a consistent gauge-invariant theory, but the bookkeeping of currents differs.  Expanding the covariant-derivative Lagrangian yields a term linear in $A_\mu$ proportional to the free Noether current, plus a quadratic term $A^2|\phi|^2$.  The conserved source current that enters the gauge-field equations is obtained by varying the full action with respect to $A_\mu$, and consequently includes the contribution of the quadratic term.  This is a standard and internally consistent feature of scalar QED.

Nevertheless, in Ref.~\cite{MinottiModanese2026GaugeCoupling} we explored a different low-energy coupling principle: the interaction is written linearly as $A_\mu J^\mu$, with $J^\mu$ the conserved Noether current of the matter sector, at the price of reducing the usual local gauge redundancy in the electromagnetic sector.  This hypothesis is motivated by the different forms taken by the bosonic and fermionic couplings and by a semiclassical interpretation of $A_\mu J^\mu$ as interaction energy.

\begingroup
As a restricted effective hypothesis, the proposal must be assessed through observables whose predictions cannot be absorbed into fitted condensate parameters. We therefore distinguish the common optical--magnetic consistency relation from the absolute-response and spectral predictions that could discriminate the couplings under additional experimental conditions.
\endgroup

\section{Standard scalar QED and the status of the conserved current}

Consider a free complex scalar field $\phi$:
\begin{equation}
\mathcal{L}_0=\partial_\mu\phi^\ast\partial^\mu\phi-m^2\phi^\ast\phi.
\end{equation}
The global phase symmetry $\phi\to e^{i\alpha}\phi$ yields the Noether current
\begin{equation}
j^\mu_{\rm free}=iq\left(\phi^\ast\partial^\mu\phi-(\partial^\mu\phi^\ast)\phi\right),
\qquad \partial_\mu j^\mu_{\rm free}=0
\quad\text{on the free equations of motion}.
\label{eq:jfree}
\end{equation}
Minimal coupling introduces $D_\mu=\partial_\mu+iqA_\mu$ and
\begin{equation}
\mathcal{L}_{\rm mat}=(D_\mu\phi)^\ast(D^\mu\phi)-m^2\phi^\ast\phi.
\label{eq:scalarQEDmat}
\end{equation}
Expanding,
\begin{equation}
(D_\mu\phi)^\ast(D^\mu\phi)=
\partial_\mu\phi^\ast\partial^\mu\phi
-iqA_\mu\left(\phi^\ast\partial^\mu\phi-(\partial^\mu\phi^\ast)\phi\right)
+q^2A_\mu A^\mu|\phi|^2.
\label{eq:expanded}
\end{equation}
The electromagnetic source is obtained by varying the full Lagrangian:
\begin{equation}
J^\mu=iq\left[\phi^\ast D^\mu\phi-(D^\mu\phi)^\ast\phi\right]
=j^\mu_{\rm free}-2q^2A^\mu|\phi|^2.
\label{eq:Jfull}
\end{equation}
This current is locally conserved as a consequence of the gauge-invariant field equations.  Hence, the standard theory contains no internal inconsistency; the coefficient of the term linear in $A_\mu$ is simply not, by itself, the complete electromagnetic source current.

\subsection{Motivation and cost of the alternative coupling}

The proposal in Ref.~\cite{MinottiModanese2026GaugeCoupling} elevates to an effective principle a form familiar from fermionic QED and classical electrodynamics: the interaction energy is linear in the potential and a conserved current.  Under $A_\mu\to A_\mu+\partial_\mu f$,
\begin{equation}
\delta\int d^4x\,A_\mu J^\mu
=-\int d^4x\,f\,\partial_\mu J^\mu,
\end{equation}
up to boundary terms.  If this term is interpreted semiclassically as the interaction energy, conservation of $J^\mu$ is a natural sufficient condition for invariance under the residual transformations allowed by the model.

\begingroup
The construction is unusual because the conserved nonrelativistic matter current used in the coupling is itself dependent on the vector potential.  For a bosonic wavefunction $\psi$ it is
\begin{equation}
\mathbf j_c=\frac{iq\hbar}{2m}\left(\psi\nabla\psi^\ast-\psi^\ast\nabla\psi\right)
-\frac{q^2}{m}|\psi|^2\mathbf A,
\qquad
\partial_t(q|\psi|^2)+\nabla\!\cdot\mathbf j_c=0.
\label{eq:jc}
\end{equation}
Thus ``linear coupling to a conserved current'' does not mean that the complete matter--field functional is linear in $\mathbf A$ after Eq.~\eqref{eq:jc} is substituted.  This specific  $\mathbf A$ dependence is precisely what produces the additional diamagnetic contribution and must be retained when the field equation is varied.
\endgroup

Reduced gauge invariance may allow an additional scalar quantity in the potentials,
\begin{equation}
S=\partial_\mu A^\mu,
\end{equation}
which in Aharonov--Bohm extended electrodynamics is sourced by $I=\partial_\mu J^\mu$.  The propagation equation, normalization, energy balance, and material response of this mode are developed in Refs.~\cite{MinottiModaneseEPJC2023,MinottiModaneseMDPI2023}; generalized local conservation in many-body descriptions is discussed in Ref.~\cite{MinottiModaneseMath2025}.  In the equilibrium transverse superconducting problem considered below, $I=0$ and $S=0$, so the scalar mode does not enter the penetration-depth solution directly.

\section{Effective-theory status and scope}

If full local gauge invariance is abandoned or restricted, the electromagnetic theory is not identical to Maxwell electrodynamics.  The proposal is not a UV-complete particle-physics theory and is not applied here to elementary-particle scattering.  In particular, the Ward identities, soft-photon factorization, KLN cancellations, and renormalizability properties of gauge-invariant QED are not assumed to survive unchanged.

\begingroup
The companion papers have distinct roles.  Ref.~\cite{MinottiModanese2026GaugeCoupling} introduces the modified coupling and derives the nonrelativistic, London, surface-impedance, and Ginzburg--Landau consequences used here.  Ref.~\cite{MinottiModaneseEPJC2023} develops gauge-wave generation and detection in the extended electromagnetic theory.  Ref.~\cite{MinottiModaneseMDPI2023} treats the scalar electromagnetic field in material media, and Ref.~\cite{MinottiModaneseMath2025} discusses generalized local conservation in many-body quantum mechanics.  The present paper derives the linked magnetic and optical responses, establishes the resulting observational degeneracy, and uses published measurements to define the requirements for consistency checks and discriminating experiments.
\endgroup

\subsection{Complementary single-particle spectral prediction}\label{sec:spectrum}

A simple single-particle consequence of the doubled diamagnetic term is obtained in a uniform magnetic field. Appendix~A derives the source-generated vector potential of a long solenoid and obtains the transverse Fock--Darwin spectrum. We take $B>0$, $\omega_c=|q|B/m$, and, for definiteness, $qB>0$; reversing the sign of $qB$ reverses the term proportional to $l$. At zero longitudinal momentum the energies are
\begin{equation}
E_{n_r,l}=\hbar\Omega(2n_r+|l|+1)-\frac{\hbar\omega_c}{2}l,
\qquad
\begingroup\Omega=\frac{\omega_c}{\sqrt{2}}.\endgroup
\label{eq:FDmain}
\end{equation}
\begingroup
For the numerical scale quoted below we take $B=10\,$T, $|q|=e$, and illustrate an ion mass $m=100m_p$:
\begin{equation}
\hbar\omega_c\simeq6.3\times10^{-9}\ {\rm eV}
\left(\frac{|q|}{e}\right)
\left(\frac{B}{10\,{\rm T}}\right)
\left(\frac{100m_p}{m}\right).
\end{equation}
For this illustrative heavy-ion case, the energy scale is extremely small. The spectral result is nevertheless conceptually complementary to penetration-depth measurements: the dependence on the angular-momentum quantum number and the lifting of the standard Landau degeneracy change the structure of the spectrum, rather than only an overall fitted condensate stiffness. Its use as an experiment would require a suitable bosonic system, controlled source geometry, and a demonstration that confinement and other interactions do not obscure the predicted level structure; no feasible implementation is established here.
\endgroup

\section{Magnetic screening and optical response from a common functional}
\label{sec:derivation}

The Ginzburg--Landau order parameter is a collective field for composite Cooper pairs.  Its parameters already contain band-structure and many-body renormalizations.  Define the conventional stiffness length
\begin{equation}
\lambda_L^{-2}=\mu_0\frac{q^2n_s}{m^\ast},
\label{eq:london}
\end{equation}
where $n_s=|\psi|^2$ in the London limit.

\begingroup
The bridge from the single-particle modified Hamiltonian to the condensate response can be written directly at Ginzburg--Landau level.  The modified matter part contains the standard covariant-gradient term plus one additional diamagnetic term,
\begin{equation}
\mathcal F_{\rm mod}=a|\psi|^2+\frac{b}{2}|\psi|^4
+\frac{1}{2m^\ast}|(-i\hbar\nabla-q\mathbf A)\psi|^2
+\frac{q^2}{2m^\ast}|\psi|^2\mathbf A^2
+\frac{\mathbf B^2}{2\mu_0},
\label{eq:Fmod}
\end{equation}
for the static transverse sector.  Here ``transverse'' means that the vector potential is divergence-free, $\nabla\!\cdot\mathbf A=0$, while the electrostatic potential and all fields are time independent.  Consequently,
\begin{equation}
S=\partial_\mu A^\mu=0,
\end{equation}
so the scalar mode of the extended electrodynamics is not excited in this configuration.  This is a property of the magnetostatic source and solution considered here, rather than an additional gauge choice; the functional in Eq.~\eqref{eq:Fmod} describes only the transverse magnetic response relevant to the penetration-depth calculation.
Varying Eq.~\eqref{eq:Fmod} with respect to $\mathbf A$ gives
\begin{equation}
\frac{1}{\mu_0}\nabla\times\mathbf B
=\frac{q\hbar}{2m^\ast i}\left(\psi^\ast\nabla\psi-\psi\nabla\psi^\ast\right)
-2\frac{q^2}{m^\ast}|\psi|^2\mathbf A.
\label{eq:Avariation}
\end{equation}
The first copy of the diamagnetic term is contained in the ordinary condensate current \eqref{eq:jc}; the second comes from the explicit extra term in Eq.~\eqref{eq:Fmod}.  For constant density and uniform phase, Eq.~\eqref{eq:Avariation} reduces to
\begin{equation}
\nabla\times\mathbf B=-\frac{2}{\lambda_L^2}\mathbf A.
\end{equation}
Taking the curl and using $\nabla\cdot\mathbf B=0$ yields
\begin{equation}
\nabla^2\mathbf B=\frac{2}{\lambda_L^2}\mathbf B,
\qquad
\lambda_{\rm mag}=\frac{\lambda_L}{\sqrt{2}}.
\label{eq:lambda_sqrt2}
\end{equation}
This reproduces the result obtained in Ref.~\cite{MinottiModanese2026GaugeCoupling} from the quasistatic potential equations.  It also makes explicit which assumption carries the single-particle modification into the many-body condensate: the same additional $q^2\mathbf A^2/(2m^\ast)$ term is included in the effective Ginzburg--Landau functional.
\endgroup

\subsection{Transverse AC response and optical spectral weight}

In the homogeneous transverse London limit, the condensate contribution to the total electromagnetic source in Eq.~\eqref{eq:Avariation} is $\mathbf J_s=-2\mathbf A/(\mu_0\lambda_L^2)$. It is the sum of the conserved matter current $\mathbf j_{c,s}=-\mathbf A/(\mu_0\lambda_L^2)$, obtained from Eq.~\eqref{eq:jc} for uniform phase, and the additional source contribution $-\mathbf A/(\mu_0\lambda_L^2)$ generated by the extra diamagnetic term. The distinction between $\mathbf j_{c,s}$ and $\mathbf J_s$ is essential: optical propagation and surface impedance are governed by the total electromagnetic source. Writing $\mathbf J_{\rm em}=\sigma_n\mathbf E+\mathbf J_s=\sigma(\omega)\mathbf E$, we use $\sigma$ for the effective transverse conductivity inferred from the electromagnetic response. With time dependence $e^{-i\omega t}$ and $\mathbf E=i\omega\mathbf A$, it is
\begin{equation}
\sigma(\omega)=\sigma_n(\omega)+\frac{2i}{\mu_0\lambda_L^2\omega}.
\label{eq:conductivity_modified}
\end{equation}
After separation of the regular contribution, this gives
\begin{equation}
\mu_0\lim_{\omega\to0}\omega\sigma_2(\omega)
=\frac{2}{\lambda_L^2},\qquad
\lambda_{\rm stiff}=\frac{\lambda_L}{\sqrt{2}}
=\lambda_{\rm mag}.
\label{eq:equal_lengths}
\end{equation}
The modified response doubles the condensate pole in $\sigma_2$ and its zero-frequency delta-function weight in $\sigma_1$. Thus a genuinely independent optical measurement of that weight still determines the modified screening length. The optical spectral-weight length is an operational response length; it must be distinguished from the parameter length $\lambda_L$ defined in Eq.~\eqref{eq:london}. This distinction is essential when comparing an optical measurement with a prediction expressed in terms of $n_s/m^\ast$. 

Including displacement current and a normal fluid with approximately constant $\sigma_n$, the bulk transverse dispersion remains
\begin{equation}
k^2=\frac{\omega^2}{c^2}+i\mu_0\sigma_n\omega-\frac{2}{\lambda_L^2}.
\label{eq:ACdispersion}
\end{equation}
In the low-frequency, low-loss limit, $k=i\kappa$ with $\kappa\simeq\sqrt{2}/\lambda_L$. The surface reactance of a local planar half-space likewise satisfies $X_s/\omega\simeq\mu_0\lambda_{\rm mag}$. The surface reactance, optical condensate pole, and static screening length therefore probe the same local transverse stiffness. At higher frequency, quasiparticle absorption, gap thresholds, and nonlocality require a material-specific conductivity rather than the simple pole alone.

\subsection{Observational degeneracy and the absolute-response prediction}

The falsifiable prediction at fixed condensate parameter is
\begin{equation}
\lambda_{\rm mag}^{-2}=2\mu_0\frac{q^2n_s}{m^\ast}
\quad\text{(effective proposal)},\qquad
\lambda_{\rm mag}^{-2}=\mu_0\frac{q^2n_s}{m^\ast}
\quad\text{(standard local model)}.
\label{eq:absolute_test}
\end{equation}
It would require an independent estimate of the relevant condensate density and effective mass, with their tensor character and interaction renormalizations controlled. Neither a penetration depth nor the low-frequency optical condensate spectral weight independently fixes this right-hand side, since each measures the same electromagnetic response kernel. A proposed microscopic density/mass calibration must be checked against its own electromagnetic model assumptions before Eq.~\eqref{eq:absolute_test} can be used. In the absence of such a calibration, a fitted $n_s/m^\ast$ can absorb the factor of two and the two theories are degenerate for the present observables.

The cancellation can be expressed more generally by writing the local condensate contribution to the electromagnetic source as $\mathbf J_s=-K\mathbf A$, with
\begin{equation}
K=\eta\frac{q^2n_s}{m^\ast},\qquad
\lambda_{\rm mag}^{-2}=\mu_0K,
\qquad
\lambda_{\rm stiff}^{-2}=\mu_0K,
\label{eq:common_kernel}
\end{equation}
where $\eta=1$ in the standard local model and $\eta=2$ in the effective proposal. Thus $R=\lambda_{\rm stiff}/\lambda_{\rm mag}=1$ is independent of $\eta$. Even exact measurements of both lengths determine $K$, but do not separately determine $\eta$ and $n_s/m^\ast$. The limitation applies to the homogeneous, local, low-frequency transverse sector considered here and does not assert equivalence of the full theories.

\subsection{Interface with the external Maxwell field}

\begingroup
Penetration depth and surface impedance are boundary-value observables.  In the present equilibrium calculation the condensate density is constant, $\nabla\cdot\mathbf A=0$, and hence $S=0$ inside the material.  In the vacuum outside, the source vanishes and the same transverse solution reduces to the Maxwell sector.  The homogeneous electromagnetic equations retain their standard form, so the normal component of $\mathbf B$ and the tangential component of $\mathbf E$ are continuous in the absence of singular magnetic or electric surface sources; the jump of tangential $\mathbf B$ is fixed by any physical sheet current. 
The additional diamagnetic term changes the bulk transverse current response and therefore both the magnetic decay constant and the optical condensate pole. It does not, by itself, modify the standard electromagnetic boundary conditions at the interface. The simple result in Eq.~\eqref{eq:lambda_sqrt2}, however, assumes a homogeneous, local bulk with a planar ideal surface. A surface layer with reduced superfluid density, a nonlocal current-response kernel, or surface roughness may produce deviations from a single exponential field profile. In that case, the penetration depth obtained from an experimental fit can contain both the intrinsic bulk decay length and surface- or material-dependent corrections. Such corrections should therefore be modeled or included among the systematic uncertainties, rather than incorporated into the bulk relation $\lambda_{\rm mag}={\lambda_L}/{\sqrt{2}}$.
\endgroup

\section{Operational definitions and measurement comparability}
\label{sec:definitions}

We use $\lambda_{\rm mag}$ for the static or quasistatic transverse screening length. Direct Meissner-state profiles are preferred; vortex-state estimates require an explicit model and field extrapolation. We use $\lambda_{\rm stiff}$ for the condensate spectral-weight length defined by
\begin{equation}
\lambda_{\rm stiff}^{-2}=\mu_0\lim_{\omega\to0}\omega\sigma_2(\omega),
\label{eq:stiffness_sigma2}
\end{equation}
after removal of the regular contribution, equivalently to the ideal condensate delta-function weight. Equations~\eqref{eq:equal_lengths} and \eqref{eq:stiffness_sigma2} imply $R\equiv\lambda_{\rm stiff}/\lambda_{\rm mag}=1$ in either local model. In a homogeneous local material, the equality is a consistency condition; failure of the equality can indicate method-dependent calibration, material inhomogeneity, nonlocal response, or missing physics, but is not the predicted signature of Eq.~\eqref{eq:Fmod}. A microwave or THz length calibrated against a magnetic measurement is not an independent absolute optical value.

\subsection{Gap scale and finite-frequency response}

\begingroup
At finite frequency, the complex conductivity depends on the ratio of $\hbar\omega$ to the superconducting gap and on impurity scattering.  In conventional superconductors the Mattis--Bardeen response describes the transfer between condensate and quasiparticle spectral weight in the local dirty limit \cite{MattisBardeen1958}.  When $\hbar\omega$ approaches the pair-breaking threshold, absorption and the reactive response cannot be represented by a frequency-independent London length.  In anisotropic or multigap materials there is no single universal threshold, and different bands can contribute differently.  Consequently, every value entering the stiffness column must be accompanied by the probing frequency range, the gap scale used in the analysis, and the model by which the condensate spectral weight was separated from the regular conductivity.
\endgroup

\subsection{Spectral-weight truncation}

\begingroup
The Ferrell--Glover--Tinkham sum rule relates the superconducting delta-function weight to the optical spectral weight removed from finite frequency \cite{FerrellGlover1958,TinkhamFerrell1959}.  Experimentally the integral is truncated at a finite cutoff and requires an assumed normal-state reference.  If recovery of the missing area extends beyond the measured window, if interband backgrounds are subtracted differently, or if the normal state evolves strongly with temperature, $\lambda_{\rm stiff}$ can be biased upward or downward.  These systematics can generate an apparent offset from the magnetic length even though both local models predict equality.  We therefore use broad ranges when the optical literature is not internally uniform and do not attach a formal statistical significance to values lacking a complete covariance or systematic-error analysis.
\endgroup

\subsection{Sample matching and material-specific effects}

\begingroup
A quantitative cross-method comparison requires recording composition, transition temperature, residual scattering or residual-resistivity information, geometry, thickness, surface preparation, crystallographic component, field, and frequency.  We define a qualitative ranking as follows (see Table \ref{tab:lambda_summary}): (A) same specimen and independent absolute determinations; (B) specimens from the same batch or closely matched composition with independently calibrated absolute values; (C) different specimens, incomplete disorder characterization, or method-dependent conversion.  Category C values illustrate the sensitivity of the consistency comparison to specimen and analysis choices. Category A would provide the strongest check of the common-kernel relation, although it would still not distinguish the two couplings by the ratio alone. However, none of the comparisons compiled in Table~\ref{tab:lambda_summary} satisfies all category-A requirements.

Impurity scattering is not a universal multiplicative correction. It can change the nonlocal-to-local crossover in Nb and Pb, act as pair breaking in superconductors with a sign-changing order parameter, and redistribute spectral weight among bands. The present comparison therefore avoids applying a single correction formula to all published values. A more controlled future consistency study would use matched specimens, report the clean or dirty regime and coherence length, and compare samples with systematically varied disorder.
\endgroup

\section{Literature comparison}
\label{sec:data}

The published comparisons below give concrete examples of the measurement distinctions defined in Sec.~\ref{sec:definitions}. They show where absolute cross-method consistency can be assessed and where calibration, specimen differences, or material response prevent a direct comparison. Values refer as closely as possible to $T\to0$. For orthorhombic YBCO, $\lambda_{ab}=\sqrt{\lambda_a\lambda_b}$ is used when both components are available.

\begingroup
\begin{table*}[t]
\caption{Penetration-depth comparison.  Lengths are in nm.  “Stiffness-side value” denotes a literature value examined as a possible estimate of the condensate stiffness from a finite-frequency or optical measurement. Some of these values are retained for completeness or sensitivity analysis even when they do not qualify as independent primary estimates. Entries marked ``ambiguous'' are not independent condensate-spectral-weight determinations under the operational definition of Sec.~\ref{sec:definitions}.  The categories are same-specimen independent absolute determinations (A), same-batch or closely matched specimens (B), and heterogeneous or method-dependent comparisons (C), as defined in Sec.~\ref{sec:definitions}.}
\label{tab:lambda_summary}
\resizebox{\textwidth}{!}{%
\begin{tabular}{lcccc}
\hline\hline
Material & stiffness-side value & $\lambda_{\rm mag}$ & ratio or range & status \\
\hline
Nb & $44$ (60 GHz, ambiguous) & $29.1\pm1.0$ & $1.51\pm0.05$ & C; sensitivity \\
Pb & $38$ (ambiguous/nonlocal) & $30.5$--$39$ & $0.97$--$1.25$ & C; excluded \\
YBCO near optimal & $126.5$--$170$ & $115.3\pm3.1$ to $118$ & $1.07$--$1.47$ & B/C \\
MgB$_2$ & $97$ & $100$ & $0.97$ & C; multiband \\
Ba(Fe,Co)$_2$As$_2$ & $300\pm30$ & $216.8\pm0.7$ & $1.38\pm0.14$ & B; different samples \\
Ba(Co,Fe)$_2$As$_2$ (Ofer) & relative 942 MHz $\Delta\lambda$ & $250\pm8$ & no independent $R$ & C; same-crystal cross-check \\
\hline\hline
\end{tabular}%
}
\end{table*}
\endgroup

\subsection{Niobium and lead}

\begingroup
The $44\,$nm Nb value of Klein \emph{et al.} was obtained from a 60 GHz surface-impedance analysis \cite{klein1994conductivity}, whereas the recent LE-$\mu$SR work gives an intrinsic London depth $29.1(10)\,$nm and explicitly analyzes nonlocal and impurity effects \cite{mcfadden2026niobium}.  Under the stricter definition above, the 60 GHz result is not automatically an independent optical spectral-weight length, so the ratio $44/29.1\simeq1.51$ is included only in the broader sensitivity analysis.  Nb, like Pb, can be nonlocal, although the modern Nb analysis finds the corrections manageable in the samples studied.  Pb remains excluded from the primary comparison because the relation among intrinsic London, Pippard, and film screening lengths is too model dependent for the present purpose \cite{Nutley1994,gasparovic1970superconducting}.
\endgroup

\subsection{YBCO}

\begingroup
Kiefl \emph{et al.} directly profiled the Meissner field in high-purity detwinned YBa$_2$Cu$_3$O$_{6.92}$ and obtained $\lambda_a=126.1\pm1.2\pm3\,$nm, $\lambda_b=105.5\pm1.0\pm3\,$nm, and $\lambda_{ab}=115.3\pm0.8\pm3\,$nm \cite{Kiefl2010YBCO}.  Their comparison table quotes an infrared result $\lambda_a=160\,$nm and $\lambda_b=100\,$nm, corresponding to $\lambda_{ab}=126.5\,$nm.  That pairing gives $R\simeq1.10$; it is a cross-method result close to unity.  Other near-optimal optical studies give values around $150$--$170\,$nm \cite{basov1995plane,homes1999effect,liu1999doping}, while magnetic analyses give approximately $117$--$118\,$nm \cite{hossain2012absolute,sonier2007}.  Pairing these heterogeneous values gives approximately $R\simeq1.07$--$1.47$; the spread illustrates the limits of cross-study comparison and does not select either coupling.  Surface dead layers of order $10\,$nm, oxygen content, chain/plane anisotropy, and optical sum-rule choices are material-specific systematics that must be resolved by same-crystal work.
\endgroup

\subsection{MgB$_2$}

\begingroup
The optical and $\mu$SR estimates, approximately $97\,$nm and $100\,$nm, are numerically consistent \cite{Seo2017MgB2Optics,Niedermayer2002MgB2muSR}.  MgB$_2$ is not treated here as a generic single-band control material.  It is a two-band superconductor, and its band-dependent gaps, scattering rates, and coherence lengths can produce non-single-component surface impedance and, in some regimes, type-1.5 phenomenology \cite{BabaevSpeight2005}.  Its values lie close to the common $R=1$ prediction, while its multiband character prevents it from serving as a simple calibration standard for single-band London electrodynamics.
\endgroup

\subsection{Ba(Fe,Co)$_2$As$_2$}

\begingroup
The comparison $300\pm30\,$nm from optical work on BaFe$_{1.85}$Co$_{0.15}$As$_2$ and $216.8\pm0.7\,$nm from a field-extrapolated $\mu$SR analysis at closely corresponding Co concentration yields $R=1.38\pm0.14$ \cite{Tu2010,Luke2010}.  The deviation from unity cannot be interpreted as evidence for either coupling: the values are from different specimens and disorder is a strong source of variation \cite{Gordon2009BaCoAs,Gordon2010AbsoluteBaCoAs,Kim2010IrradiatedBaCoAs}.

Ofer \emph{et al.} provide a useful same-crystal comparison: LE-$\mu$SR gives an absolute $\lambda(0)=250(8)\,$nm, and 942 MHz microwave cavity perturbation gives the relative temperature variation $\Delta\lambda(T)$ on a piece of the same crystal \cite{Ofer2012BaCoAs}.  The microwave curve was shifted to overlap the absolute $\mu$SR scale, so the two methods give a consistent temperature dependence within that procedure, but the microwave result is not an independent absolute determination of the numerator in $R$.  It therefore constrains method-dependent discrepancies without constituting a direct $R=1$ measurement.
\endgroup

\begingroup
\begin{figure}[t]
\centering
\begin{tikzpicture}
\begin{axis}[
width=0.94\columnwidth,
height=0.72\columnwidth,
xlabel={$\lambda_{\rm mag}$ (nm)},
ylabel={stiffness-side length (nm)},
xmin=20,xmax=270,ymin=20,ymax=350,
legend style={font=\scriptsize,at={(0.03,0.97)},anchor=north west},
label style={font=\small},tick label style={font=\scriptsize},
]
\addplot[domain=20:270,samples=2] {x};
\addlegendentry{$R=1$}
\addplot+[only marks,mark=*,error bars/.cd,
x dir=both,x explicit,y dir=both,y explicit]
coordinates {
(116.7,148.25) +- (1.7,21.75)
(100,97) +- (0,0)
(216.8,300) +- (0.7,30)
};
\addlegendentry{MgB$_2$, YBCO, Ba(Fe,Co)$_2$As$_2$}

\addplot+[only marks,mark=o,error bars/.cd,
x dir=both,x explicit]
coordinates {(29.1,44) +- (1.0,0)};
\addlegendentry{Nb, method ambiguous}

\end{axis}
\end{tikzpicture}
\caption{Graphical cross-method comparison; the $R=1$ line is predicted by both local models.  The filled points
represent, from lower to higher magnetic penetration depth, MgB$_2$,
near-optimally doped YBCO, and Ba(Fe,Co)$_2$As$_2$.  Their classifications
in Table~\ref{tab:lambda_summary} are respectively C, B/C, and B.  The open
point represents Nb, whose 60 GHz stiffness-side value is classified as
methodologically ambiguous and is included only in the sensitivity
analysis.  The vertical error bar for YBCO represents the range of
literature values; this range does not represent a formal error bar. The
limited number of points and the dependence on measurement classification
and material-specific systematics are apparent.}
\label{fig:comparison}
\end{figure}
\endgroup

\section{Cross-method sensitivity and interpretation}
\label{sec:stats}

The published values do not support a conventional meta-analysis: several lack complete uncertainties, the points are not statistically homogeneous, and systematic errors may be correlated. A descriptive residual from the common local prediction can be written as
\begin{equation}
D(1)=\left[\frac{1}{N}\sum_{i=1}^N(R_i-1)^2\right]^{1/2}.
\end{equation}
For the representative central ratios $R_i=(1.51,1.27,0.97,1.38)$, including the method-ambiguous Nb point, $D(1)\simeq0.35$. Excluding Nb gives $D(1)\simeq0.27$. These numbers are descriptive diagnostics, not uncertainties, significance levels, or measures of relative support for standard and modified coupling. Choosing the lower infrared YBCO estimate moves that material closer to unity. The selection and classification of measurements dominate the comparison.

\section{Experimental implications and discriminating observables}
\label{sec:experiments}

\subsection{A controlled optical--magnetic consistency check}

An informative first stage is to determine the absolute Meissner profile and condensate spectral weight independently on the same specimen, for the same tensor component and temperature. Composition, surface condition, disorder, gap scales, the regular optical contribution, and the spectral cutoff must be controlled. Such a category-A comparison would test whether one local transverse kernel describes both measurements. It would also help locate discrepancies associated with optical weight extraction, surface layers, or nonlocal screening. Agreement with $R=1$ is a consistency result shared by both couplings; disagreement would require analysis of these effects or of physics beyond the assumed local sector.

The literature assessment identifies practical priorities for this stage. Independent absolute optical calibration is essential even when relative temperature dependences agree. Same-specimen work removes an important ambiguity in the YBCO and Ba(Fe,Co)$_2$As$_2$ comparisons, while Nb and Pb require explicit treatment of nonlocality and the distinction between intrinsic and fitted screening lengths. Multiband materials require a response analysis that resolves the contributions relevant to the measured tensor component.

\subsection{An absolute-response test with independent condensate information}

A test of Eq.~\eqref{eq:absolute_test} needs a separately justified absolute value of $q^2n_s/m^\ast$ that does not simply re-express the optical or magnetic kernel. In a composite condensate, a band mass, total carrier density, or normal-state spectral weight cannot automatically be identified with the required condensate parameter: interaction renormalizations, condensate participation, and tensor character must be accounted for within an explicit microscopic description. An independent mass estimate alone is sufficient only if the relevant condensate density is also constrained. The complete uncertainty and model dependence must be small enough to resolve the factor-of-two difference in inverse-square screening length.

This requirement is the essential distinction between measuring the response more accurately and identifying its coupling coefficient. Without independent condensate information, improved optical and magnetic precision constrains $K$ in Eq.~\eqref{eq:common_kernel}, but leaves its decomposition into coupling and condensate parameters undetermined.

\subsection{A complementary spectral route}

The spectrum in Sec.~\ref{sec:spectrum} and Appendix~A provides a different type of prediction. In the specified source-generated potential, the additional $\mathbf A^2$ term removes the guiding-center degeneracy of the standard Landau problem and introduces a distinct angular-momentum dependence. Resolving the level pattern, rather than a single overall energy scale, could in principle supply information beyond a fitted stiffness. This route remains conditional on a system for which the effective Hamiltonian applies, the potential is fixed by a controlled source, and finite-size and interaction effects can be separated. The heavy-ion estimate given above illustrates a significant sensitivity limitation rather than an experimental proposal ready for implementation.

\section{Conclusions}

The effective Ginzburg--Landau functional studied here gives a common modification of magnetic screening and optical condensate response. At fixed $q^2n_s/m^\ast$, both inverse-square penetration lengths are doubled, and therefore $\lambda_{\rm mag}=\lambda_{\rm stiff}=\lambda_L/\sqrt{2}$. The standard local theory also predicts equality of the two operational lengths. Their ratio consequently eliminates the coupling coefficient and cannot distinguish these theories within the homogeneous local transverse approximation.

This non-discrimination result specifies what penetration-depth comparisons can establish. Independent magnetic and optical measurements test the consistency of the transverse response and of its experimental extraction. Our literature assessment makes this role concrete by separating absolute from relative or externally calibrated determinations, classifying specimen matching, and identifying material-specific limitations. The available comparisons include useful consistency checks but no fully independent same-specimen category-A determination; their scatter does not constitute evidence for either coupling.

Discriminating tests require information beyond the ratio. The absolute-response prediction becomes informative only when the relevant condensate density and effective mass are justified independently of the response being tested. The modified single-particle spectrum offers a complementary structural prediction, subject to substantial system-selection and sensitivity requirements. Together, these results define a reasonable experimental strategy: establish cross-method consistency under controlled conditions, then seek independent condensate information or observables with a different dependence on the modified coupling.

\section*{Data availability}
No new data were generated in this study. All data analyzed are available in the published sources cited in the article.

\appendix

\section{Spectrum of an integer-spin particle in a uniform magnetic field}

In ordinary electromagnetism it is possible to impose the Lorenz gauge condition, which for static configurations with time-independent scalar potential reduces to $\nabla\cdot\mathbf A=0$.  In the extended Aharonov--Bohm electrodynamics the four-potential is instead determined by its source equation.  In the magnetostatic regime of a long solenoid,
\begin{equation}
\nabla^2\mathbf A(\mathbf r)=-\mu_0\mathbf J(\mathbf r).
\label{eq:poissonA}
\end{equation}
For an ideal infinitely long solenoid of radius $R$, carrying azimuthal surface current $\mathbf K=K\hat{\boldsymbol\varphi}$, a source-generated solution is
\begin{equation}
A_\varphi(r)=
\begin{cases}
\dfrac{B}{2}r,&r<R,\\[5pt]
\dfrac{BR^2}{2r},&r>R,
\end{cases}
\qquad B=\mu_0K.
\label{eq:AphiSolenoid}
\end{equation}
Inside the solenoid,
\begin{equation}
\mathbf A=\frac12\mathbf B\times\mathbf r=\frac{B}{2}(-y,x,0),
\qquad \mathbf B=B\hat z.
\label{eq:AsymBulk}
\end{equation}
This is the symmetric-gauge form, but here it is obtained as the interior limit of the specified source.

The modified bosonic Hamiltonian is
\begin{equation}
H_{\rm AB,bos}=\frac{1}{2m}(\mathbf p-q\mathbf A)^2+\frac{q^2}{2m}\mathbf A^2.
\label{eq:HABbos}
\end{equation}
For the transverse motion, take $p_z=0$, $B>0$, and $qB>0$, and define $\omega_c=|q|B/m$. Using $\mathbf A^2=B^2r^2/4$,
\begin{equation}
H_{\rm AB,bos}=\frac{\mathbf p^2}{2m}
+\frac12m\Omega^2r^2-\frac{\omega_c}{2}L_z,
\qquad
\begingroup\Omega=\frac{\omega_c}{\sqrt{2}}.\endgroup
\label{eq:HABbosFD}
\end{equation}
The eigenvalues are
\begin{equation}
E_{n_r,l}=\hbar\Omega(2n_r+|l|+1)-\frac{\hbar\omega_c}{2}l,
\quad n_r=0,1,2,\ldots,\quad l\in\mathbb Z.
\end{equation}
For nonzero longitudinal momentum, add the free-motion contribution $p_z^2/(2m)$. For $qB<0$, the sign of the term proportional to $L_z$, and hence of the term proportional to $l$ in the energies, is reversed. Unlike the standard Landau problem, the transverse energy depends on $l$ and the guiding-center degeneracy is removed. The uniform-field approximation requires the transverse extent of the state to be much smaller than the solenoid radius $R$. For low-lying states a sufficient scale separation is $R\gg\ell_B$, with
\begin{equation}
\ell_B=\sqrt{\frac{\hbar}{|q|B}}.
\end{equation}

\bibliographystyle{ieeetr}
\bibliography{references_revision_merged}

\end{document}